\documentclass[%
 reprint,
superscriptaddress,
 amsmath,amssymb,
 aps,
pra,
]{revtex4-2}
\usepackage{graphicx}
\usepackage{mhchem}
\usepackage{comment}
\usepackage[normalem]{ulem}
\usepackage{graphicx}
\usepackage{dcolumn}
\usepackage{bm}
\usepackage{hyperref}
\usepackage{soul}
\usepackage[dvipsnames]{xcolor}
\setstcolor{red}

\begin{document}
\title{Memory-Driven Self-Propulsion and Flocking of Chemically Active Droplets}

\author{Samuel Kovach}
\affiliation{Department of Physics and Center for Biomolecular Condensates, Washington University in St. Louis, St. Louis, MO 63130 USA}

\author{Trevor GrandPre}
\email{trevorg@wustl.edu}
\affiliation{Department of Physics and Center for Biomolecular Condensates, Washington University in St. Louis, St. Louis, MO 63130 USA}
\affiliation{National Institute for Theory and Mathematics in Biology, Northwestern University and The University of Chicago, Chicago, IL, USA}

\date{\today}

\begin{abstract}
Biomolecular condensates are continually remodeled by biochemical reactions
that can exhibit non-Markovian, history-dependent dynamics. We develop a theory
of active phase separation with non-Markovian reactions and show that delayed
reaction feedback destabilizes stationary droplets: when the memory time
becomes comparable to the reaction turnover time, condensates deform and
spontaneously acquire a polar, self-propelled state. In multidroplet systems,
persistent memory wakes mediate alignment, producing polar flocks and, at
higher concentrations, traveling labyrinths. These results establish reaction
memory as a control parameter of active phase separation, linking condensate
remodeling, autonomous motility, and collective organization, and suggest a
possible route to flocking-like behavior within cells.
\end{abstract}

\maketitle

Biomolecular condensates—intracellular compartments formed via phase separation—operate out of equilibrium.  Active processes regulate their formation, dynamics, and function. In particular, cytoskeletal components such as microtubules and actin filaments, along with molecular motors and chemical reactions, introduce persistent molecular fluxes that regulate condensate number, size, and dynamics. Experiments and theoretical work have begun to uncover how these nonequilibrium processes shape the physical behavior and biological function of condensates~\cite{aierken2026roadmap}.

A growing body of experimental work shows that active processes give condensates
material properties inaccessible at equilibrium: they can
enhance or suppress phase separation, control condensate size, and drive
oscillatory dynamics and
fluidization~\cite{patel2017atp,dine2018protein,he2026kinase,sastre2025size}.
Activity also produces collective behavior---self-propulsion, collective
organization, and
flocking~\cite{ashraf2025emergence,hokmabad2021emergence,dwivedi2025chemical}---and
accelerates and regulates redox and enzymatic
reactions~\cite{smokers2024droplets,yu2026origins,dai2023interface,papp2025biomolecular,stoffel2025enhancement}.
Active condensates further participate in cellular mechanotransduction, sensing
and relaying mechanical
forces~\cite{case2015integration,sawada2006force,swaminathan2016fak,case2022synergistic,ghosal2026principles,hu2014fak}.

From theory and computer simulations, a diverse set of nonequilibrium behaviors in biomolecular condensates has begun to emerge. Coarse-grained and bead-per-residue simulations demonstrate how active processes and sequence-dependent interactions regulate condensate structure and dynamics~\cite{wani2026non}. Active and reaction-driven models have demonstrated phenomena including accelerated and reverse Ostwald ripening, arrested coarsening, oscillatory dynamics, and droplet size selection~\cite{glotzer1995reaction,zwicker2015suppression,wurtz2018stress,zwicker2025physics,wurtz2018chemical,sorkin2025accelerated,brauns2021wavelength,seyboldt2018role,bauermann2025theory, zippo2025molecular,laha2024chemical, heltberg2026oscillating, prathyusha2024anomalous, haugerud2026excitability}. These systems can also exhibit microphase separation and finite-wavelength pattern formation~\cite{tjhung2018cluster,fries2025active,saha2020scalar,brauns2024nonreciprocal,martis2025trait, bauermann2025critical}, as well as directed or self-propelled condensate motion~\cite{demarchi2023enzyme,saha2020scalar,schede2023model,you2020nonreciprocity, sorkin2026propelling, saha2025effervescence, qiang2025self}. Recent work has shown that memory in the mobility of conserved dynamics can
destabilize diffusive phase separation, producing traveling waves and
self-organized spatiotemporal structures even in the absence of chemical
reactions~\cite{gajendragad2025memory}. Here we focus on reaction memory in
active condensates---history-dependent, non-Markovian reaction kinetics---which
remains largely unexplored in continuum theories of phase separation. We develop a continuum theory that couples these delayed reactions to diffusive
transport and phase separation, and show that reaction memory drives the
emergence of motile and collective phases.

There are several biological mechanisms through which non-Markovian reaction rates may arise~\cite{teza2025coarse, van2022thermodynamic, blom2024milestoning, igoshin2025coarse}. For example, signaling cascades involving sequences of intermediate reactions can introduce effective delays, leading to memory effects in the dynamics \cite{zhang2019markovian}. Similarly, a multivalent protein with multiple phosphorylation sites populates many states of differing valency that continually interconvert. The condensate therefore incorporates a mixture of these states whose composition evolves in time. Another important example
involves chemically active proteins, such as kinases, that switch between sticky
and non-sticky states and often colocalize with other chemically active
proteins~\cite{zippo2025molecular,he2026kinase}. The resulting combinations of
molecular states can be described as a multi-state system within condensates,
which, upon coarse-graining, yields an effective two-state description with
history-dependent transition rates. Moreover, biochemical reactions are generally non-instantaneous: they involve finite residence times in intermediate states as well as transition times between states, both of which can introduce memory into the effective dynamics~\cite{chung2018transition}.
It remains unclear how such history-dependent reactions influence the mechanics and dynamics of condensates. Here, we develop a theory of chemically active droplets with temporally nonlocal reaction kinetics. This provides a framework to study how reaction memory shapes condensate behavior.

\begin{figure*}[t]
    \centering
    \includegraphics[width=\textwidth]{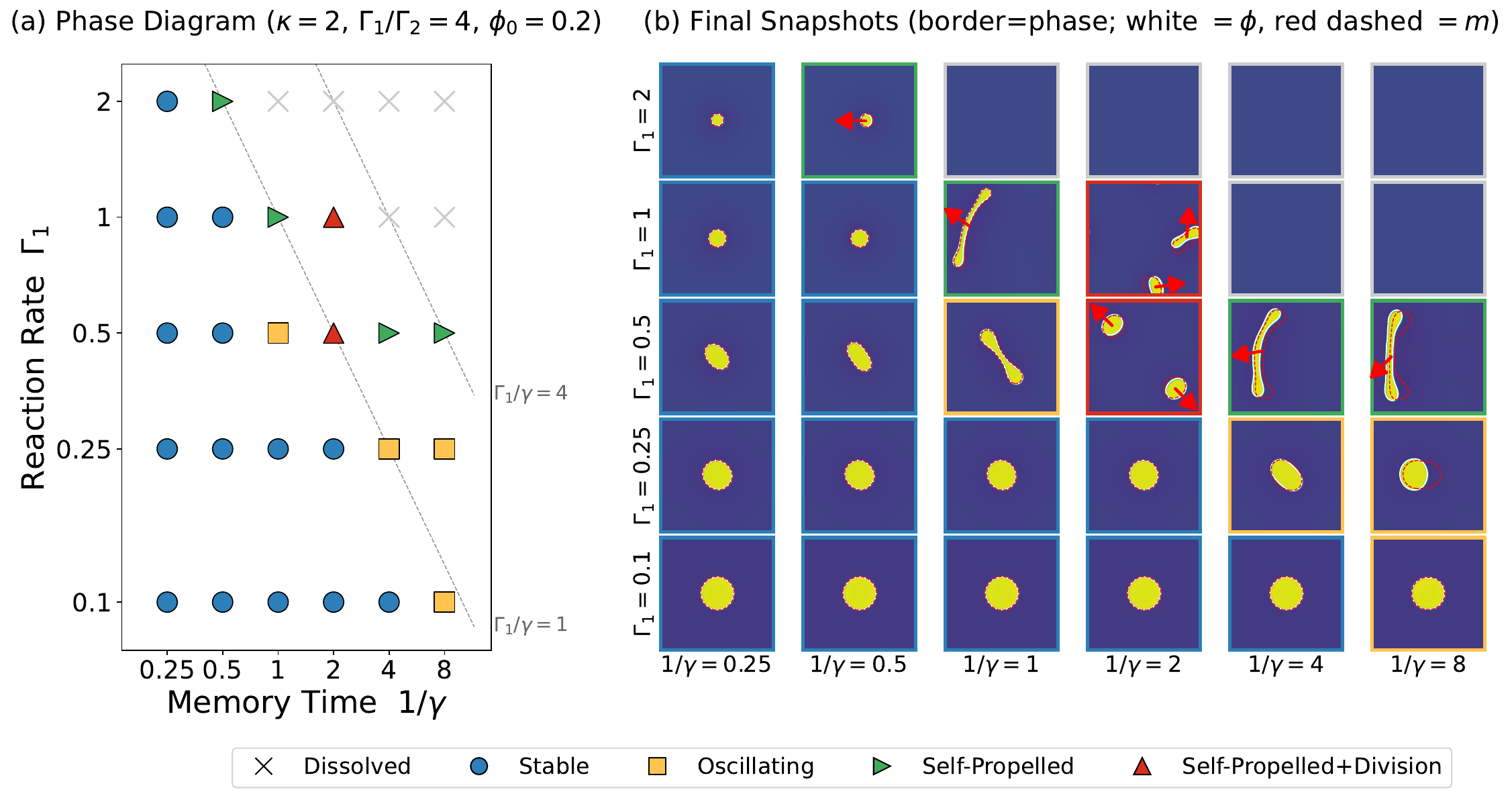}
    \caption{
    Phase behavior of memory-driven active condensates.
    \textbf{(a)} Phase diagram as a function of the memory time
    $1/\gamma$ and reaction rate $\Gamma_1$ for
    $\kappa=2$, $\Gamma_1/\Gamma_2=4$, and $\phi_0=0.2$.
    Symbols indicate the observed dynamical states.
    Dashed lines show constant values of $\Gamma_1/\gamma$.
    \textbf{(b)} Representative final snapshots corresponding to the
    parameter values shown in panel (a). Colored borders indicate the
    classified phase, the white region denotes the condensate field
    $\phi$, and the red dashed contour denotes the memory field $m$ when it has decayed to a value of $1/e$.
    }
    \label{fig:memory-phase-diagram}
\end{figure*}
One biologically motivated class of active condensates is driven by dissipative
chemical reactions, minimally modeled as two components---a phase-separating
species (A) and a non-phase-separating species (B)---interconverted by reactions
that break detailed balance. These dynamics generically produce nonequilibrium steady states beyond a Boltzmann description, leading to finite droplet size selection controlled by reaction rates~\cite{glotzer1995reaction,zwicker2015suppression}. In most existing studies, these rates are taken to be constant; however, in biological settings, they are expected to depend on both space and time within condensates. These dependencies may lead to important consequences for both transient dynamics and steady-state behavior. Previous work has shown that chemically active droplets can evade classical Ostwald ripening, attain a stable finite size, and even undergo cell-like growth and division through reaction-driven nonequilibrium turnover~\cite{zwicker2017growth}. Here we show that temporal memory in the reaction kinetics enriches this
phenomenology and provides a generic route to self-propulsion: shape
fluctuations generate delayed reaction fields that break fore--aft symmetry and
set droplets into motion.

Self-propelled droplets have been reported in several other reaction–diffusion and chemically active systems. For example, enzyme-enriched condensates can self-propel through catalytic feedback between substrate/product gradients and enzyme fluxes~\cite{demarchi2023enzyme}, chemically active mixtures can self-propel through nontransitive phase coexistence~\cite{qiang2025self}, and active phase separation can propel catalytic colloidal structures~\cite{sorkin2026propelling}. More broadly, chemically active droplets are known to swim through Marangoni or phoretic mechanisms driven by self-generated solute gradients~\cite{maass2016swimming,michelin2023self}, and phase-separated droplets can migrate along dissolution-favoring chemical gradients~\cite{izri2014self}. Our mechanism differs from these examples because propulsion is generated by temporal memory in the reaction kinetics. The delayed reaction field becomes misaligned with the instantaneous droplet shape. It will therefore be important to systematically map the phases of the model for a given memory kernel.

\begin{figure}[hbt!]
    \centering
    \includegraphics[width=\columnwidth]{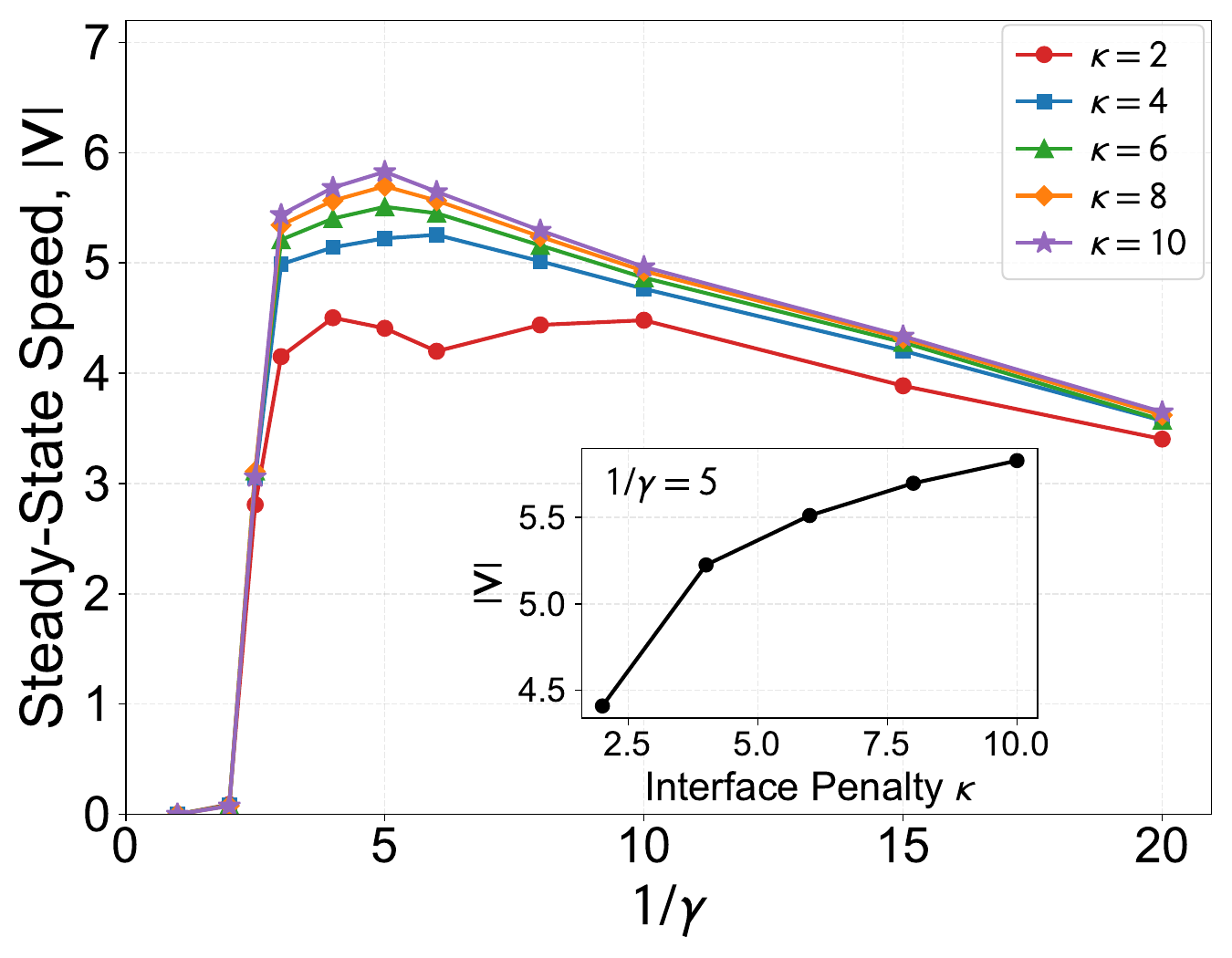}
    \caption{
    Tunable propulsion speed of a memory-driven active condensate.
    Steady-state speed $|\mathbf{V}|$ versus memory time $1/\gamma$ for
    interface penalties $\kappa=2$--$10$ (fixed $\Gamma_1=0.5$,
    $\Gamma_1/\Gamma_2=4$, hence $\alpha=\Gamma_1+\Gamma_2=0.625$; $\chi=2.5$).
    The speed is nonmonotonic: it vanishes below an onset near
    $1/\gamma\approx2$, peaks at intermediate memory $1/\gamma\approx5$, and
    decays at long memory. The onset is $\kappa$-independent---all curves rise
    together---and is thus set by the reaction rate $\alpha$ rather than by the
    interfacial stiffness, whereas the peak speed grows monotonically with
    $\kappa$ (inset, at $1/\gamma=5$): the reactions set the onset and the
    interfacial stiffness sets the magnitude. Speeds are in simulation units
    (grid length per unit time); $|\mathbf{V}|$ is the Savitzky--Golay time
    derivative~\cite{savitzky1964smoothing} of the $\phi$-weighted condensate
    centroid, taken as the median over the second half of each run.
    }
    \label{fig:speed-sweep}
\end{figure}

Within this work, we consider non-Markovian chemical reactions of the form
\begin{equation}
\label{reactions}
\ce{A <=>[{\Gamma_{AB}(t-t')}][{\Gamma_{BA}(t-t')}] B}
\end{equation}
where $A$ represents one or more molecular states that participate in phase
separation, while $B$ represents one or more molecular states that do not. The rates $\Gamma_{AB}(t-t')$ and $\Gamma_{BA}(t-t')$ encode history-dependent transitions from state $A$ to $B$ and vice versa. Within this framework, $t$ would be the current time and $t^{\prime}$ is some time in the past. Our goal is to demonstrate the physical consequences enabled by history-dependent rates.

The reactions in Eq.~\ref{reactions} are coupled to a Cahn–Hilliard reaction–diffusion model:
\begin{equation}
\label{model1}
\partial_t \phi(\mathbf{r},t)
=
M(\phi)\nabla^2 \frac{\delta F}{\delta \phi}
+
s(\phi,t),
\end{equation}
where $M(\phi)$ is the concentration-dependent mobility, the reaction term is given by

\begin{equation}
\label{model2}
\begin{split}
s(\phi,t)
=
\int_{0}^t dt^{\prime}
\Big[
&
\Gamma_{BA}(t-t^{\prime})
(1-\phi(\mathbf{r},t^{\prime}))
\\
&
-
\Gamma_{AB}(t-t^{\prime})
\,\phi(\mathbf{r},t^{\prime})
\Big]~,
\end{split}
\end{equation}
and the free energy takes the form
\begin{equation}
F
=
\frac{k_B T}{v}
\int d^d r
\left[
f(\phi)
+
\frac{\kappa}{2}
|\nabla \phi|^2
\right].
\end{equation}
Here, $\kappa$, $d$, $k_B$, $T$, and $v$ denote: the interfacial energy penalty; the dimensionality of the system; the Boltzmann constant; the temperature; and the coarse graining volume; respectively. More generally, one could replace the linear dependence on \(\phi\) by a nonlinear reaction functional \(\mathcal R[\phi]\) in Eq.~\eqref{model2}. Here we focus on the minimal linear two-state model. The results do not depend deeply on the type of free energy, but for the sake of clarity, we consider a Flory-Huggins free energy $f(\phi)=\phi\ln \phi+(1-\phi)\ln (1-\phi)+\chi\phi(1-\phi)$. The field $\phi$ is the packing fraction of the phase separating component A, and $\chi$ is the energetic term for interaction.

Figure~\ref{fig:memory-phase-diagram} shows the phase diagram in
$(1/\gamma,\Gamma_1)$ at $\kappa=2$, $\Gamma_1/\Gamma_2=4$ (fixing the
homogeneous steady state $\phi_0=0.2$), and $\chi=2.5$ (See Appendix~\ref{numerics} for simulation details). We write each reaction kernel as an effective rate times a
normalized temporal distribution,
\begin{equation}
\begin{aligned}
    \Gamma_{AB}(\tau)&=\Gamma_1\,\hat\Gamma_{AB}(\tau),\\
    \Gamma_{BA}(\tau)&=\Gamma_2\,\hat\Gamma_{BA}(\tau),
\end{aligned}
\end{equation}
with $\hat\Gamma_{BA}$ and $\hat\Gamma_{AB}$ normalized,
$\int_0^\infty\hat\Gamma_{BA}\,d\tau=\int_0^\infty\hat\Gamma_{AB}\,d\tau=1$. Thus $\Gamma_1$ and $\Gamma_2$ are the integrated forward and backward rates, and
the total relaxation rate is
\begin{equation}
    \alpha=\Gamma_1+\Gamma_2=\int_0^\infty \Delta(\tau)\,d\tau ,
\end{equation}
where $\Delta(\tau)=\Gamma_{AB}(\tau)+\Gamma_{BA}(\tau)$ is the combined memory
kernel. For the phase diagram we take exponential
profiles $\hat\Gamma_i(\tau)=\gamma e^{-\gamma\tau}$, for which the reactions
act on a memory field $m$ obeying $\partial_t m=\gamma(\phi-m)$---an
exponentially weighted history of $\phi$ with memory time $1/\gamma$.

For short memory the droplets remain stable
[Fig.~\ref{fig:memory-phase-diagram}(b)]. As the memory time approaches the
droplet relaxation time, oscillatory shape instabilities emerge and drive
spontaneous propulsion, forming an intermediate self-propelled band at
$\Gamma_1/\gamma\sim1$. At larger amplitudes the droplets divide, and the
daughters self-propel apart. For large reaction rates or long memory, the
reactions overwhelm phase separation and the condensate dissolves.

First, we show the behavior at small memory, $\gamma> \Gamma_1$. Linearizing about the uniform steady state $\phi_0$,
$\phi(\mathbf r,t)=\phi_0+\delta\phi(\mathbf r,t)$, and Fourier transforming in
space, $\delta\phi_{\mathbf k}(t)=\int d^dr\,e^{-i\mathbf k\cdot\mathbf r}
\delta\phi(\mathbf r,t)$, Eqs.~\ref{model1}--\ref{model2} give
\begin{equation}
\label{linear1}
\partial_t\delta\phi_{\mathbf k}(t)
= D(k)\,\delta\phi_{\mathbf k}(t)
-\int_0^t d\tau\,\Delta(\tau)\,\delta\phi_{\mathbf k}(t-\tau),
\end{equation}
with $k=|\mathbf k|$. Here
$D(k)=2\kappa\Lambda k^2(k_c^2-k^2)$ is the Cahn--Hilliard dispersion
relation~\cite{glotzer1995reaction}, with $\Lambda=M(\phi_0)k_BT/v$ and
$k_c^2=|f''(\phi_0)|/2\kappa$. In the two-phase region $f''(\phi_0)<0$, so
$D(k)>0$ for $k<k_c$ and the homogeneous state is unstable to phase
separation. We keep only the
leading interfacial contribution and neglect terms that remain $O(1)$ at
nonzero $k$.

\begin{figure*}[t]
    \centering
    \includegraphics[width=\textwidth]{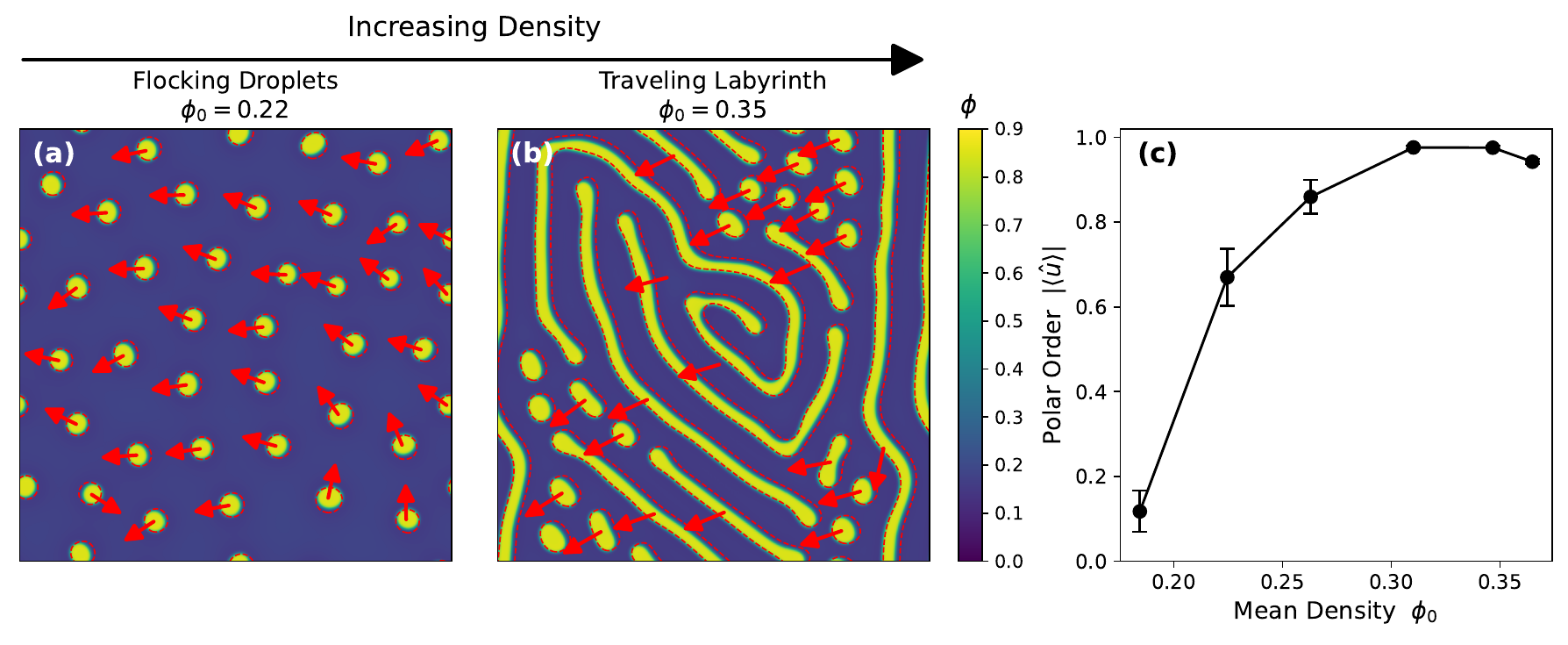}
    \caption{
    Collective motion of memory-driven active condensates with increasing density.
\textbf{(a,b)} Late-time snapshots ($t=60$) of the condensate field $\phi$,
evolved from multi-droplet initial conditions at mean density $\phi_0=0.22$ (a)
and $\phi_0=0.34$ (b), for $\kappa=2$, $\chi=2.5$, $\Gamma_1=2$, and memory time
$1/\gamma=0.5$; the reaction ratio $\Gamma_1/\Gamma_2$ is chosen such that the
steady-state fraction equals $\phi_0$, so the reactions preserve the mean
density. Red arrows indicate the self-propulsion direction
$\hat{\mathbf{u}}_i$ of each condensate, given by the dipole moment of the
memory lag,
$\hat{\mathbf{u}}_i \propto \int_{A_i}\left[\phi(\mathbf{r})-m(\mathbf{r})\right]
(\mathbf{r}-\mathbf{r}_i)\,\mathrm{d}^2r$,
where $A_i$ is the region occupied by condensate $i$ and $\mathbf{r}_i$ its
centroid: the memory field trails a moving condensate, so $\phi-m>0$ on its
leading edge. At low density the droplets remain discrete and align into a
polar flock (a); at high density they merge into bands traveling transverse to
their long axis (b).
\textbf{(c)} Polar order parameter
$\left|\langle\hat{\mathbf{u}}\rangle\right|
=\bigl|\tfrac{1}{N}\sum_{i=1}^{N}\hat{\mathbf{u}}_i\bigr|$,
where $N$ is the number of condensates, as a function of the mean density
$\phi_0$; values are time-averaged over the final $10$ time units and error
bars denote the standard deviation over that window.
$\left|\langle\hat{\mathbf{u}}\rangle\right|\to1$ corresponds to complete
alignment.
    }
    \label{fig:flocking-density}
\end{figure*}
In steady state, the field becomes time independent and the memory kernel decouples from the concentration field. The reaction term reduces to $s_{\mathbf{k}} = -\alpha,\delta\phi_{\mathbf{k}}$, so that the steady-state exterior concentration field satisfies a screened Poisson equation
\begin{equation}
    \left(k^2+\ell^{-2}\right)\delta\phi_{\mathbf{k}} = 0,
\end{equation}
where $\ell^2=D_{\rm eff}/\alpha$ sets the reaction-screening length $\ell$,
with effective diffusivity $D_{\rm eff}=2\kappa\Lambda k_c^2$.

In this stationary regime, the droplet-size theory reduces to the results established in previous work, with the non-Markovian reaction dynamics entering only through the integrated rates $\Gamma_1$, $\Gamma_2$, and $\alpha$~\cite{zwicker2015suppression}. Deviations from this steady-state description arise when the long-time dynamics remain time dependent: the frequency-dependent memory kernel then stays coupled to the concentration field at all times, and no separation of timescales exists.

To characterize the dynamical transitions, we examine how the linear growth
rate depends on memory.  Assuming a
normal-mode perturbation
$\delta\phi_{\mathbf{k}}(t)=\delta\phi_{\mathbf{k}}(0)\,e^{\omega(k)t}$ and
substituting into Eq.~\eqref{linear1} gives the self-consistent dispersion
relation
\begin{equation}
\label{gr}
\begin{aligned}
\omega(k)&=D(k)-\Delta(\omega),\\
\Delta(\omega)&=\int_0^\infty d\tau\,\Delta(\tau)\,e^{-\omega\tau}.
\end{aligned}
\end{equation}
The memory correction $\Delta(\omega)$ is the Laplace transform of the kernel,
so Eq.~\eqref{gr} is implicit in $\omega(k)$ and must be solved
self-consistently, unlike the Markovian case.

For the exponential kernel $\Delta(\tau)=\alpha\gamma\,e^{-\gamma\tau}$ the
transform is a single pole, $\Delta(\omega)=\alpha\gamma/(\omega+\gamma)$, and
Eq.~\eqref{gr} becomes quadratic,
\begin{equation}
\label{quad}
\omega^2+\bigl[\gamma-D(k)\bigr]\omega+\gamma\bigl[\alpha-D(k)\bigr]=0,
\end{equation}
with roots
\begin{equation}
\label{omega_exp}
\omega(k)=\tfrac{1}{2}\Bigl[\bigl(D(k)-\gamma\bigr)
\pm\sqrt{\bigl(D(k)+\gamma\bigr)^2-4\gamma\alpha}\,\Bigr].
\end{equation}
The finite memory time $1/\gamma$ thus replaces the single Markovian branch
$\omega=D(k)-\alpha$ with two: the extra root is the relaxation mode of the
memory field, now an independent degree of freedom.

Both roots have negative real part for $D(k)<\min(\alpha,\gamma)$, so
instability sets in at whichever of $\alpha,\gamma$ is smaller. For
$\alpha<\gamma$, it occurs at $D(k)=\alpha$, where Eq.~\eqref{quad} factorizes
as $\omega(\omega+\gamma-\alpha)=0$: a stationary instability ($\omega=0$, with
a decaying mode $\omega=\alpha-\gamma<0$). For $\gamma<\alpha$, it occurs first
at $D(k)=\gamma$, where the roots form a conjugate pair
$\omega_\pm=\pm i\Omega_{\rm H}$ with $\Omega_{\rm H}=\sqrt{\gamma(\alpha-\gamma)}$
that gains a positive real part for $D(k)>\gamma$. Slow memory relaxation
($\gamma<\alpha$, i.e.\ memory time exceeding the reaction time $1/\alpha$) thus
converts the stationary size-control instability into a growing Hopf
oscillation, initiating the self-propelling dynamics. In Fig.~\ref{fig:memory-phase-diagram}(a), $\alpha=1.25\,\Gamma_1$ (since
$\Gamma_2=\Gamma_1/4$), so the $\Gamma_1/\gamma$ guide lines closely track the
$\alpha/\gamma=1$ onset, and the measured instability thresholds agree
quantitatively with the linear theory.

The droplet velocity is tunable through the reaction rates and interfacial
stiffness $\kappa$, and is nonmonotonic in the memory time, peaking at an
intermediate $1/\gamma$ (Fig.~\ref{fig:speed-sweep}). Propulsion is generated
by the lag of the memory field behind the concentration. For a droplet moving
at velocity $\mathbf{V}$ the memory lags as
$m\simeq\phi+\gamma^{-1}\mathbf{V}\cdot\nabla m$, so
$\phi-m\simeq-\gamma^{-1}\mathbf{V}\cdot\nabla m$ is a dipole aligned with the
motion---positive on the leading edge and negative in the trailing wake (red
contours, Fig.~\ref{fig:memory-phase-diagram})---which reinforces the displacement
and sustains propulsion. The speed vanishes in both limits: for short memory
the lag is negligible, and for long memory the memory field is nearly frozen
and decouples from the droplet; it is largest when the memory time is
comparable to the droplet's relaxation time.

Finally, we ask how multiple droplets interact. Starting from a multidroplet
initial condition [Fig.~\ref{fig:flocking-density}; $\kappa=2$, $\chi=2.5$,
$\Gamma_1=2$, $1/\gamma=0.5$, with the reaction ratio set so that the
homogeneous steady state equals the mean density $\phi_0$], we find that the
self-propelled droplets align cooperatively, with the degree of alignment
growing with $\phi_0$. The droplets couple through the shared memory field:
each trails a memory wake [red dashed contours in
Fig.~\ref{fig:flocking-density}(a,b)], and a droplet entering the wake of a
neighbor ahead feels an asymmetric reaction field that reorients its
propulsion, consistent with wake-mediated alignment. As $\phi_0$ increases, this coupling
strengthens and the polar order $|\langle\hat{\mathbf u}\rangle|$ rises toward
unity [Fig.~\ref{fig:flocking-density}(c)]. At low density the droplets remain
discrete and form a polar flock [Fig.~\ref{fig:flocking-density}(a)]; at high
density they merge into an interconnected, collectively translating labyrinth
of stripes and droplets [Fig.~\ref{fig:flocking-density}(b)].

In summary, temporal memory in reaction kinetics transforms stationary
size-controlled condensates into oscillating, deforming, and self-propelled
droplets, with speeds tunable by the reaction rates and interfacial stiffness. In multidroplet systems the same delayed
reaction field couples droplets through persistent memory wakes, yielding polar
alignment and, at higher density, traveling bands. Our minimal
Cahn--Hilliard reaction--diffusion model isolates this mechanism; adding
hydrodynamics or richer kernels may shift thresholds while
preserving the underlying delayed feedback. Whether phosphorylation cycles,
enzyme residence times, or hidden biochemical states generate comparable
motility \emph{in vitro} and in cells remains open. More broadly, memory emerges
as a distinct control parameter of active phase separation---a general route
from condensate remodeling to autonomous motion and collective organization.

 \section{Appendix}

\subsection{Numerical Calculation}
\label{numerics}
For our numerical study of the droplets, we utilized the Dedalus software package \cite{2020PhRvR...2b3068B}. Dedalus uses the spectral method to solve partial differential equations. For this work, we evolved Eq. \ref{model1} on a $256 \times 256$ grid for Fig. \ref{fig:memory-phase-diagram}; and a $128 \times 128$ grid for Figs. \ref{fig:speed-sweep} and \ref{fig:flocking-density}. Regardless of grid size, the simulations were always ran with periodic boundary conditions. To evolve our system, we used the 2nd-order semi-implicit BDF scheme \cite{5f76054b-e77e-3a4a-93f6-21c4274a03aa} with a time step of $1\times10^{-4}$. For the studies involving a singular droplet—Figs. \ref{fig:memory-phase-diagram} and \ref{fig:speed-sweep}—the droplet was initialized in the center of the grid. The initial radius of the droplet was set near that of the stable value. A hyberbolic tangent function was used to soften the boundary between the interior and exterior of the droplet. In our calculation where many droplets were considered, we initialized the droplets one of two ways. In the case of the flocking droplets—Fig. \ref{fig:flocking-density}(a)—the droplets were initialized dilute and in random positions. As for the traveling labyrinth—Fig. \ref{fig:flocking-density}(b)—the droplets were initialized into a slab of droplets in the middle of the box.

In order to evolve the system, the following convolution between the exponential memory kernel and the concentration field needed to be solved:
\begin{equation}
    m(t)=\gamma \int_{-\infty}^{t}dt^\prime \;e^{-\gamma(t-t')}\phi(t').
\end{equation}
We evaluate $m$ by constructing an ODE through differentiating $m$ with respect to $t$. By using Leibniz's rule we find,
\begin{equation}
    \frac{dm}{dt}=\gamma(\phi-m).
\end{equation}
We then step our system in the following manner; 
\begin{equation}
    m(t+ dt)=e^{-\gamma dt}m(t)+(1-e^{-\gamma dt})\phi(t).
\end{equation}
This form reduces to $m(t+ dt)=(1-\gamma d t)m(t)+\gamma d t\,\phi(t)$ when $\gamma dt$ is small. For our initial memory field we used a grid entirely composed of the homogeneous steady state value $\phi_0$. 
\begin{acknowledgments}
We thank Gianluca Teza, Luca Cocconi, and Alex Holehouse for useful discussions. This research was supported in part by grant NSF PHY-2309135 to the Kavli Institute for Theoretical Physics (KITP). T.G. acknowledges support from the National Institute for Theory and Mathematics in Biology through the National Science Foundation (grant number DMS-2235451) and the Simons Foundation (grant number MP-TMPS-00005320).
\end{acknowledgments}

\end{document}